\documentclass[aps,twocolumn,groupedaddress,showpacs,nofootinbib]{revtex4}

\usepackage{graphicx}
\usepackage{hyperref}
\usepackage{dcolumn}
\usepackage{bm}
\usepackage{epsfig,amsbsy,bm,amssymb}
\usepackage{xcolor}

\renewcommand{\k}{{k}}
\newcommand{\hs}{\hspace*}
\newcommand{\vs}{\vspace*}
\newcommand{\np}{\newpage}
\newcommand{\w}{\omega}
\newcommand{\W}{\Omega}

\newcommand{\eref}[1] {(\ref{#1})}
\newcommand{\Eref}[1] {Eq.~(\ref{#1})}
\newcommand{\Fref}[1] {Fig. \ref{#1}}

\newcommand{\nn}{\nonumber}
\newcommand{\be}{\begin{equation}}
\newcommand{\ee}{\end{equation}}
\newcommand{\br}{\begin{eqnarray*}}
\newcommand{\er}{\end{eqnarray*}}
\newcommand{\ba}{\begin{eqnarray}}
\newcommand{\ea}{\end{eqnarray}}
\newcommand{\bp}{\begin{minipage}}
\newcommand{\ep}{\end{minipage}}
\newcommand{\bt}{\begin{tabular}}
\newcommand{\et}{\end{tabular}}

\newcommand{\bs}{\bigskip}
\newcommand{\ms}{\vspace*{-5mm}}
\newcommand{\mms}{\vspace*{-2.5mm}}
\renewcommand{\l}{\lambda}

\renewcommand{\k}{{\bm k}}

\newcommand{\x}{{\bm x}}
\newcommand{\y}{{\bm y}}

  \newcommand{\A}{{\bm A}}

\renewcommand{\t}{\tau}
\renewcommand{\l}{\lambda}

\newcommand{\Wcm}[2]{
$\rm {#1}\times10^{{#2}}~W/cm^2$}

\newcommand{\ask}[1]{\textcolor{blue}{#1}\hs{-1mm} }
\renewcommand{\ask}[1]{\textcolor{black}{#1}\hs{-1mm} }

\begin{document}
\bibliographystyle{apsrev}
\title{Random shots phase retrieval by angular streaking of XUV
  ionization}

\title{Ionization phase retrieval by angular streaking from random shots of XUV
  radiation}

\author{A.~S. Kheifets$^{1}$}
\author{R. Wielian$^{1}$}
\author{V.~V. Serov$^{2}$}
\author{I.~A. Ivanov$^{3}$}
\author{A. Li Wang$^{4}$}
\author{A. Marinelli$^{4}$}
\author{J.~P. Cryan$^{4}$}

\affiliation{$^{1}$Research School of Physics, The Australian National
  University, Canberra ACT 2601, Australia} 
\email{A.Kheifets@anu.edu.au}

\affiliation{$^{2}$General, Theoretical and Computer
  Physics, Saratov State University, Saratov
  410012, Russia}

\affiliation{$^{3}$Center for Relativistic Laser Science, Institute
  for Basic Science (IBS), Gwangju 61005, South Korea}

\affiliation{$^4$Stanford PULSE Institute,
SLAC National Accelerator Laboratory,  Menlo Park, CA 94025 USA}

 \date{\today}


\begin{abstract}
We demonstrate an accurate phase retrieval of XUV atomic ionization by
streaking the photoelectron in a circularly polarized IR laser field.
Our demonstration is based on a numerical solution of the
time-dependent Schr\"odinger equation.  We test this technique using
the hydrogen atom ionized by isolated attosecond XUV pulses across a
wide range of the photon energies. Importantly, the proposed method
works in a random shot mode when the time delay between the ionizing
and streaking XUV and IR pulses may vary from shot to shot. This is a
significant advantage over the existing interferometric techniques
which require a systematic and controllable scan of the XUV/IR delay
in one set of measurements. Such a scan may not be possible to take
because of the arrival time jitter inherent to the stochastic nature
of self-amplified FEL radiation.
 \end{abstract}

\maketitle

\ask{\section{Introduction}}

Angular streaking of extreme ultraviolet (XUV) atomic ionization with
a circularly polarized IR laser radiation has become a useful tool for
characterizing isolated attosecond pulses (IAP) from free-electron
laser (FEL) sources.  First suggested theoretically
\cite{Zhao2005,Kazansky2016,SiqiLi2018,Kazansky2019}, this method has
been implemented in practice for a shot-to-shot characterization of
IAP at FEL \cite{Hartmann2018,Duris2020}. Angular streaking of XUV
ionization (ASXUVI or ASX for brevity) combines elements of the
two previously developed techniques: attosecond angular streaking
known as the attoclock \cite{Eckle2008,Eckle2008nphys,Pfeiffer2012}
and the attosecond streak camera (ASC)
\cite{Constant1997,ItataniPRL2002,Goulielmakis2004,Kienberger2004,Yakovlev2005,Fruhling2009,Zhang2011,PhysRevLett.107.213605}. It
is analogous to ASC in that XUV pulses are
the primary source of ionization while it is common to the attoclock
in that the ionized electrons interact with a circularly polarized
laser field which rotates the photoelectron momentum distribution
(PMD). This rotation is most graphic in the plane
perpendicular to the laser propagation direction.  The difference with
the attoclock is that the latter is a self-referencing technique where
both tunneling ionization and steering are driven by the same
elliptical infrared (IR) laser pulse.  In its original form
\cite{Kazansky2016,SiqiLi2018,Kazansky2019,Hartmann2018,Duris2020},
ASX was considered in an intense IR laser field and analyzed
within the strong field approximation (SFA) \cite{XiZhao2022}. An
alternative view within the lowest order perturbation theory (LOPT)
\cite{0953-4075-45-18-183001,Dahlstrom201353,Maquet2014} considers IR
streaking as an interference phenomenon which opens a natural access
to the streaking phase $\Phi_S$. The latter is typically decomposed
into the XUV ionization  (or Wigner) phase and the laser induced
(or continuum-continuum - CC) phase. These two
phases can be converted to the corresponding time delay components,
which add up to the atomic time delay.  This delay is often
interpreted as the time it takes for the electron to be photoionized
plus the time it takes for the measurement process to occur. This
timing analysis has been used to determine the attosecond time delay
in photoemission from atoms \cite{M.Schultze06252010} and solid
surfaces \cite{Cavalieri2007}. The latter seminal works opened up the
rapidly growing field of attosecond chronoscopy of photoemission
\cite{C3FD00004D,RevModPhys.87.765,Thumm2015}. 

A straightforward extension of time-resolved ionization studies to
novel FEL sources of XUV radiation is not possible at present.  The
existing interferometric techniques rely on a systematic and
controllable scan of the XUV/IR pulse delay in one set of
measurements.  The stochastic nature of self-amplified FEL radiation
and the inherent timing jitter do not permit such a scan. In the
meantime, ASX allows for the retrieval of the XUV ionization phase and
the associated timing information from a series of XUV shots arriving
at the target atom randomly.
In this Letter, we demonstrate this useful phase retrieval capability
of ASX over a wide range of photoelectron energies right from the
ionization threshold and exceeding it many times. We demonstrate the
accuracy of the proposed technique by considering the hydrogen atom
driven by a femtosecond XUV pulse across a wide range of carrier
frequencies.  Our demonstration is based on a numerical solution of
the time-dependent Schr\"odinger equation (TDSE). It is verified
through comparison with the well established RABBITT (reconstruction
of attosecond beating by interference of two-photon transitions)
technique. Both methods are shown to provide \ask{comparable} phase
and timing information.

\begin{figure}[h]
\bp{5cm}
\epsfxsize=4.5cm
\epsffile{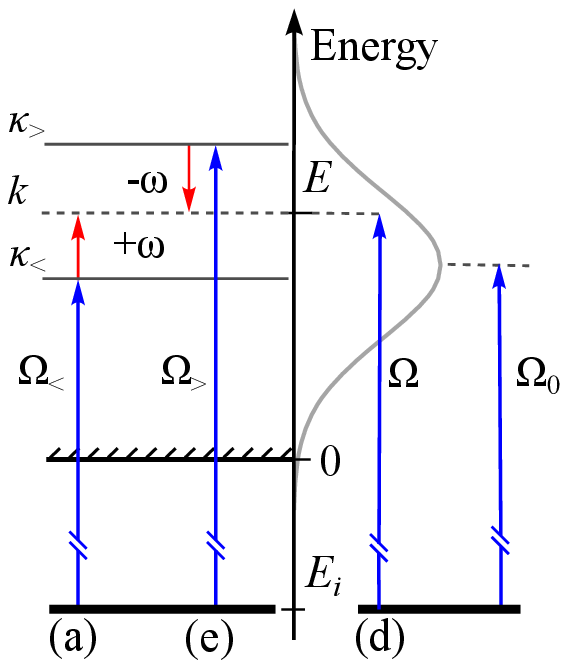}
\ep
\hs{-0.75cm}
\bp{5cm}
\includegraphics[scale=1.2,angle=270]{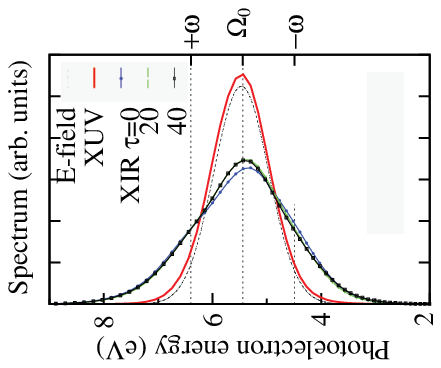}
\ep

\caption{Left: Schematic representation of IR streaking of XUV
  ionization. The first order direct process (d) and the second order
  processes aided by an IR photon absorption (a) and emission (e) are
  labeled accordingly (adopted from \cite{Dahlstrom201353}). Right:
  photoelectron spectra of H driven by XUV and XIR are
  overlapped with the  Fourier transform of the electric field.
\label{Fig1}}
\end{figure}

\np
\vs{-2cm}
\ask{\section{Physical interpretation}}
\ms

An interference character of IR streaking of XUV ionization is
illustrated in the left panel of \Fref{Fig1}. There are three
ionization channels marked (d), (a) and (e) leading to the same final
state with the photoelectron energy $E$. While the direct channel (d)
contains an unassisted XUV ionization, two other channels are aided by
an IR absorption (a) or emission (e). For these three channels to
interfere, the spectral width of the XUV pulse should be large enough
to accommodate the $\pm\w$ laser assisted processes. In the time
domain, this means that the XUV pulse is shorter than the laser
period.  The spectrum of such a pulse is illustrated in the right
panel of \Fref{Fig1} and corresponds to ionization of the hydrogen
atom at the central XUV and IR photon energies $\W_0=0.7$~au and
$\w=0.038$~au, respectively, while FWHM of the XUV pulse is set to
2~fs. The photoelectron spectrum of an unassisted XUV photoionization
(process d) is an exact replica of the Fourier transform of the
electric $E$-field shifted by the ionization potential.  Meanwhile, the
photoelectron spectrum of XUV+IR ionization (XIR for brevity) is
significantly broadened by the second-order (a) and (e) processes.

The proposed phase retrieval  by ASX is based on the following
consideration. We apply the SFA and write the photoionization
amplitude as \cite{Kitzler2002}
\be
    b(\k ,\tau) = i \int_{t_0}^{\infty} \!\! dt \ E_x (t-\tau) 
 D_x\left[\k -\A (t)\right]  e^{-i\Phi(t)}
\ .
\ee
Here the electric field of the XUV pulse $E_x$ is advancing the
streaking pulse by the time $\t$. The streaking field is described by
its vector potential 
$$
\A (t) = 
A_0\cos(\omega t)\hat{\x} +A_0\sin(\omega t)\hat{\y}
\ ,
$$ 
while the photoelectron momentum is confined
to the polarization plane  $\k = k\cos\phi~\hat{\x} +
k\sin\phi~\hat{\y}$, where $\phi$ is the emission angle. 

\begin{figure}[t]
\vs{-0.5cm}
\hs{-2mm}
\epsfxsize=8.5cm
\epsffile{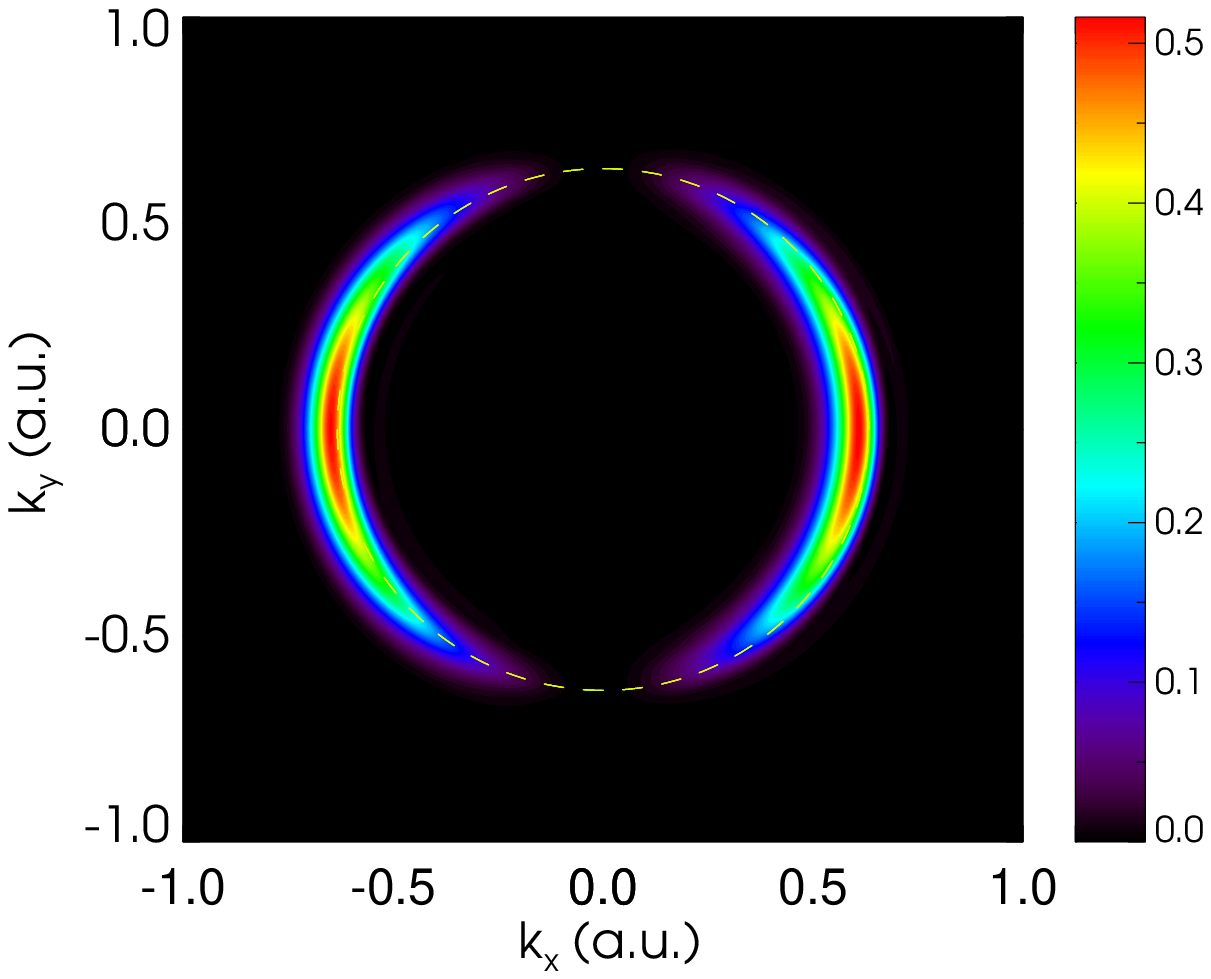}

\vs{-1.2cm}
\epsfxsize=8.5cm
\epsffile{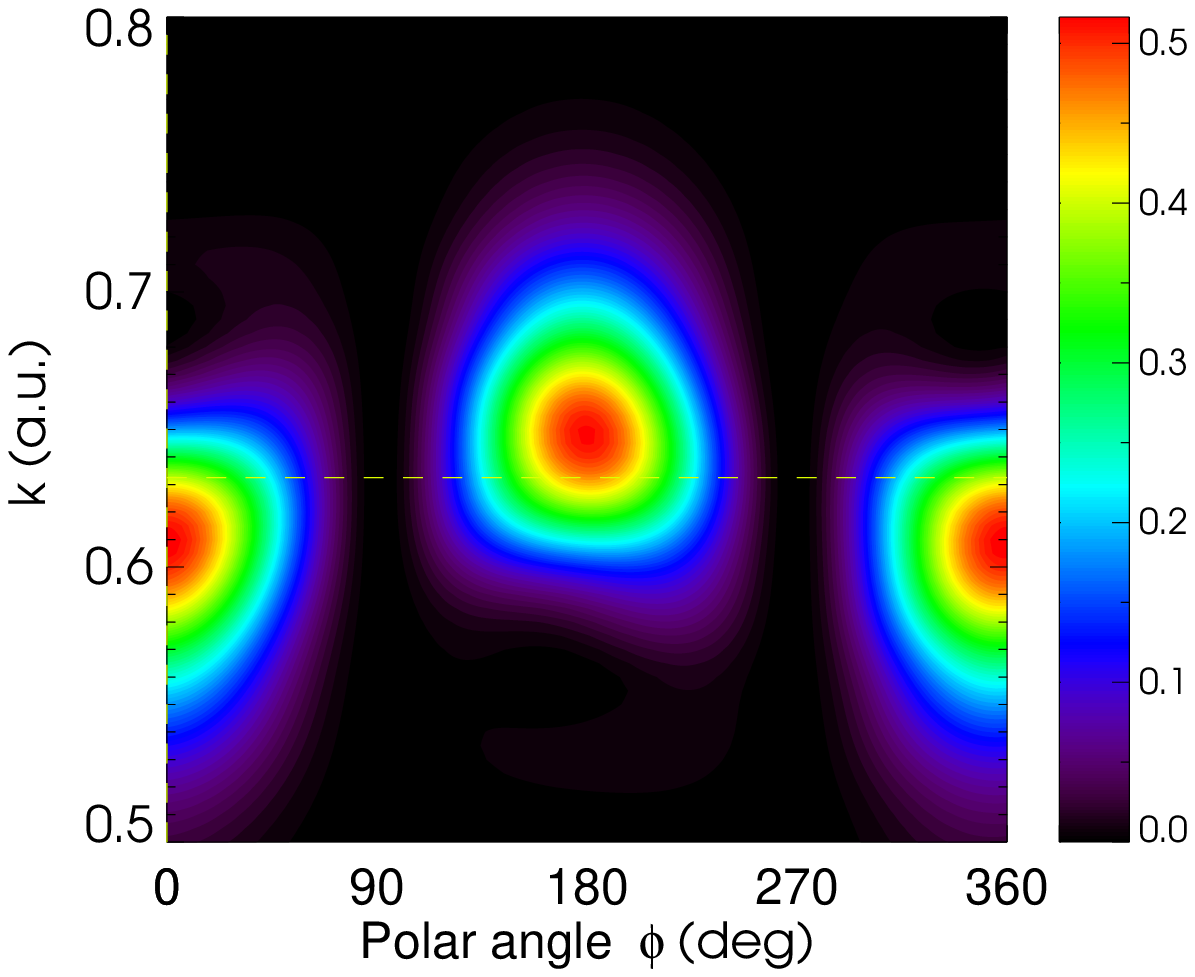}
\ms\ms

\caption{The PMD of the hydrogen atom in the polarization plane
  displayed in the Cartesian) and polar (bottom) coordinates. The
  dashed circle (top) and the horizontal straight line (bottom)
  display the momentum value $k_0=\sqrt{2(\W_0-I_p)}$ from the energy
  conservation.  The solid vertical lines (bottom) define the angular
  integration limits when calculating the radial momentum profiles
  $P_\pm(k,\t)$. The XUV/IR time delay $\tau=0$ in this figure.  \ms
\label{Fig2}}
\end{figure}

The exponential term contains the so-called Volkov phase
\be
\Phi(t)=
\frac12\left[\int_{t}^{\infty} dt^{\prime} 
\left[\k  - \A (t^{\prime})\right]^2 - k_0^2 \, t\right] 
\ ,
\ee
in which  the photoelectron energy in the absence of streaking
 $E_0=k_0^2/2=\W-I_p$. 
This phase can be estimated by the saddle point method (SPM) which
selects the most probable photoelectron trajectories leaving the atom at the time
$t_{\rm st}$ and keeping the phase stationary:
\be
\label{SFA} 
\Phi'(t_{\rm st}) = \frac12 |\k-\A(t_{\rm st})|^2-E_0=0 
\ee 
We assume that the XUV pulse is short relative to the IR pulse and
shifted relative to its peak position by the time $\t$. Under these
conditions, \Eref{SFA} is transformed to the following {\em~isochrone}
equation \cite{Kazansky2016}:
\be
\label{iso}
k^2/2-E_0 = kA_0\cos(\phi-\w\t) 
 ; \  
\left\{
\begin{array}{rl}
k>k_0 & \phi-\w\t\sim0\\
k<k_0 & \phi-\w\t\sim180^\circ
\end{array}
\right.
\ee
Here we neglected the ponderomotive energy $U_p=A_0^2/2$ in a weak
streaking field.  
The bottom panel of \Fref{Fig2} illustrates the isochrone equation
\eref{iso} most straightforwardly. The PMD extrema, which correspond
to $\phi=0$ and $180^\circ$ at $\t=0$, are displaced relative to the
photoelectron momentum $k_0$ (dashed line) by the amount $\pm A_0$.

In the above derivation it was implicitly assumed
that the phase of the dipole matrix element $D\left[\k -\A (t)\right]$
does not depend on the photoelectron energy. In a more general case,
this phase contains an energy dependent term
\be
\arg  \left\{D\left[\k -\A (t)\right]\right\} 
\propto \alpha |\k-\A (t)|^2/2
\ ,
\ee
where
\vs{-0.5cm}
\be
\label{alpha}
\alpha = \partial \arg D(\sqrt{2E})/ \partial E
\ee
is a generalized definition of the photoelectron group delay in
photoemission (see e.g. Eq.~(S10) of \citet{M.Schultze06252010}). The
introduction of this delay modifies the stationary phase equation:
\be 
\frac12 \left|\k - \A (t_{st})\right|^2 - E_0 +
\frac{\alpha}{2}\frac{d}{dt}\left[ \left(\k - \A (t_{\rm st})\right)^2 \right]
= 0 
\ee
This  leads to a modified isochrone equation
\begin{eqnarray}
\label{isom}
k^2/2 - E_0 &=& kA_0 
\left[ \cos(\phi - \w \t) 
- \alpha\w\sin(\phi - \w \t) \right]
 \nn \\ &\approx & kA_0
\cos[\phi - \w \t +  \w\alpha ]
\end{eqnarray}
In the second  line, we used the  identity
%
$  b\cos x - a\sin x
= \sqrt{1+(ab)^2} \cos(x+y) 
 , \ y=\tan^{-1}(a/b)
$
and  the long wavelength approximation $\w\alpha\ll 1$ for the 
streaking field. 
Thus the isochrone acquires an additional phase shift
$\Phi_S=\w\alpha$ which can be extracted from the photoelectron
momentum distribution (PMD).  In the following, we shall demonstrate
that $\alpha=\Phi_S/\w=\t_a$ under certain XUV and IR pulse
parameters.  \ask{For simplicity of derivation, $\alpha$ was
  introduced in \Eref{alpha} solely as the dipole phase derivative. In
  practice, we will show that it also contains the measurement
  induced CC components.}

\bs

\ask{\section{Numerical results}}

\begin{figure}[t]
\ms

\epsfxsize=7cm
\epsffile{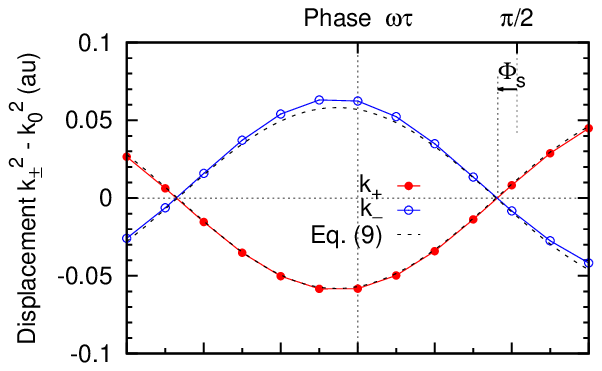}
\ms\mms

\epsfxsize=7cm
\hs{-5mm}
\epsffile{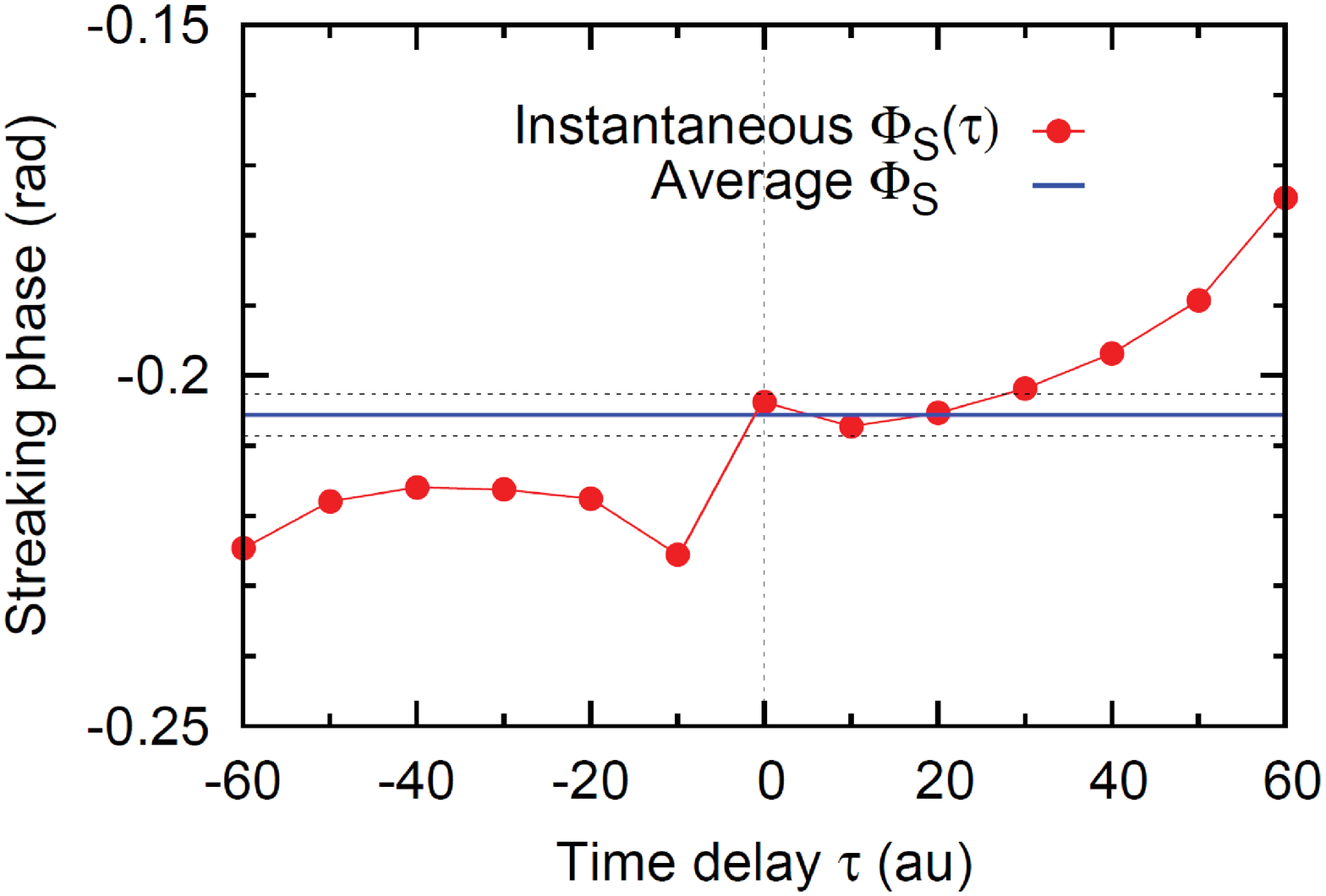}

\mms
\caption{Top: radial momentum displacements $k_\pm^2/2-k_0^2/2$ are
  shown at various XUV/IR delays $\t$. The dashed line represents the
  fit with \Eref{ansatz}. The arrow indicates the streaking phase
  $\Phi_S$.  Bottom: the fit with \Eref{ansatz} is applied to
    individual $\t$ values to determine the instantaneous
    $\Phi_S(\t)$. The average $\Phi_s$ is shown as a solid line with
    error bars visualized by dotted lines. 
\label{Fig3}}
\ms
\end{figure}

In \Fref{Fig2} we display the PMD of the hydrogen atom ionized with a
2~fs XUV pulse at \Wcm{6}{13} and $\W_0=0.7$~au, in the presence
  of a long IR pulse with $\w=0.038$~au ($\l=1200$~nm), FWHM = 25~fs  at
\Wcm{1.5}{11}. The XUV pulse is linearly polarized along the $\hat\x$
axis whereas the IR pulse is circularly polarized in the $(xy)$
plane. The PMD is projected on this plane. 
The top panel illustrates the dipole angular pattern $\propto
\cos^2\phi$ of the primary XUV ionization of the hydrogen $1s$ initial
state which is largely retained in the XIR ionization.  The bottom
panel shows the polar coordinate representation
highlighting the momentum offset induced by the vector potential of
the laser pulse. In the exhibited case, the XUV/IR delay $\tau=0$.
To simulate XIR ionization numerically, we run a series of TDSE
calculations using the computer code \cite{PhysRevA.84.062701}. At
each XUV photon energy, we scan the delay between the XUV pulse and
the IR laser field ($\tau$) in the range of 0 to 70 au in 8
increments.

The modified isochrone equation \eref{isom} allows, in principle, for
the ASX phase extraction at each pair of the variables $k,\phi$. For
better clarity, it is more advantageous to conduct such an extraction at
the extrema of the PMD. When a limited statistics is available
experimentally, the whole lobe of the PMD centered at these extrema
can be used. We demonstrate this by integrating the PMD over the $\pm
90^\circ$ angular intervals centered around $\phi=0$ and $\pm
180^\circ$ (shown by vertical  lines in the bottom panel of
\Fref{Fig2}) . We carry out this procedure at several fixed increments of
the XUV/IR delay~$\tau$. The corresponding radial momentum profiles
$P_+(k,\t)$ are shown in the top panel of \Fref{Fig3} for the angular
integration range $|\phi|<90^\circ$ The analogous profiles $P_-(k,\t)$
corresponding to the $|180^\circ -\phi|<90^\circ$  integration
range mirror closely the set shown in the figure.  The two sets of the
momentum profiles allow to determine the mean photoelectron momenta at
each $\t$
$$
k_\pm(\t) = \int kP_\pm(k,\t) dk \Big / \int P_\pm(k,\t) dk
\ ,
$$
which are then used to obtain the isochrone phase offset:
\be
\label{ansatz}
k_\pm^2(\t)/2-E_0 = \pm A_0\, k_\pm(\t)\cos(\w\t+\Phi_S)
\ .
\ee
In our simulation, the ansatz \eref{ansatz} can be applied to the
whole set of the time delays $\tau$ treating $\Phi_s$ as a common
fitting parameter. This way an average $\Phi_S$ value is obtained as
illustrated in the top panel of {Fig3}. Alternatively, we can solve
\Eref{ansatz} for each $\t$ value and to obtain an instantaneous
$\Phi_S(\t)$ value. This value would mimic a streaking phase retrieved
in a single shot measurement at a given $\t$. Both sets of streaking
phases are shown in the bottom panel of \Fref{Fig3} where the average
$\Phi_S$ is displayed with the straight solid line while its error
bars are exhibited with the dotted lines. While the deviation of
$\Phi_S(\t)$ from $\Phi_S$ is exceeding the error bars of the latter,
the method works quite satisfactorily even if a single $\tau$
determination of the streaking phase is made. The attained accuracy of
about 20\% is on par with that achieved in other streaking measurements
(see e.g. \cite{SchmidPRL2019}).

We note that \Eref{ansatz} contains the vector potential magnitude
which may not be known experimentally because of the laser focus
averaging effect. After a sufficient number of XUV shots at random
$\t$ is collected, an effective vector potential magnitude can be
determined from the maximum vertical displacement between the upwards
and downwards shifted lobes $A_0=0.5|k_+-k_-|_{\max}$ across the whole
series. Statistical nature of such a determination will not be
detrimental to the accuracy of the phase retrieval which is rather
insensitive to the IR pulse intensity.  For instance, at
  $\W=0.7$~au and the IR pulse intensity of \Wcm{1.5}{11}, the
  streaking phase $\Phi_s=-0.196$~rad. This value changes to
  $-0.177$~rad at \Wcm{1.2}{11} and to $-0.211$~rad at
  \Wcm{1.8}{11}. While these two intensities differ by 50\%, the
  streaking phase $\Phi_s$ varies by only 10\%.  The streaking phase
is similarly insensitive to ellipticity. The cited value of 
$\Phi_s=-0.196$~rad  changes only in the fourth significant figure when
the circular polarization is reduced to an elliptical one with
$\epsilon=0.9$. At a sufficiently large XUV photon energy, the
streaking phase tends to zero as a fast photoelectron is represented
by a plane wave. Hence the maximum vertical displacement $2A_0$ is
attained at $\t=0$. This provides a convenient calibration of the XUV
pulse arrival time.

\begin{figure}[t]
\epsfxsize=9cm
\epsffile{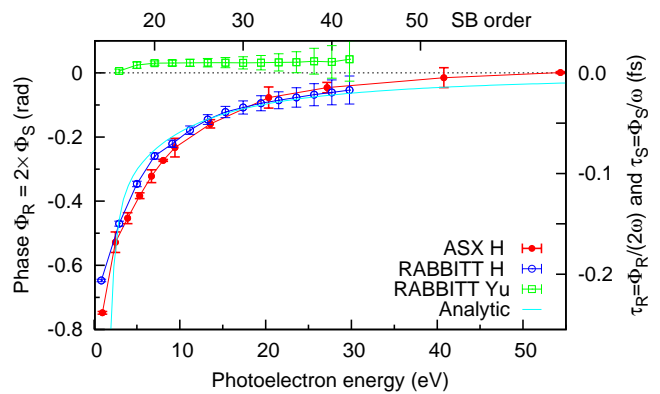}

\caption{Left axis:  RABBITT phase $\Phi_R$ as extracted from
  \Eref{RABBITT} is compared with the twice the streaking phase
  $2\Phi_s$ obtained from the isochrone ansatz \Eref{iso}.
Right axis:  \ask{RABBITT time delay $\t_R=\Phi_R/(2\w)$ and the streaking
time delay $\tau_S=\Phi_S/\w$} are compared with the Coulombic Wigner time
delay augmented by the CC correction from
\cite{Pazourek2013,Serov2015b}. The RABBITT phase and time delay in
the Yukawa atom (Yu) are also shown.
\label{Fig4}}
\ms
\end{figure}

The avarage streaking phases $\Phi_S$ at various XUV photon energies
are displayed in \Fref{Fig4}. These phases are benchmarked against our
present RABBITT calculations on hydrogen at the corresponding XUV energies and
the same IR intensity and wavelength.  Numerical
details of these calculations can be found in preceding works
\cite{PhysRevA.97.063404,PhysRevA.104.L021103}.  From these
simulations, we obtain the RABBITT phase $\Phi_R$ by fitting the
angular integrated photoelectron spectra at various $\t$ with the
following expression:
\ba
\label{RABBITT}
S_{2q}(\tau) &=&
A+B\cos[2\omega\tau-\Phi_R(E)]
\ea
Here the sideband index $2q$ defines the photoelectron energy
$E=2q\w-I_p$. We note that the amplitudes of the RABBITT peaks
\eref{RABBITT} oscillate with twice the laser frequency $\propto
2\w\t$ and the atomic time delay is defined as $\t_R=\Phi_R/(2\w)$.
In the meantime the streaking signal \eref{ansatz} contains the
$\propto \w\t$ oscillation and we expect the atomic time delay to be
expressed as $\t_S=\Phi_S/\w$. Therefore, we make a comparison of the
RABBITT phase $\Phi_R$ with twice the streaking phase $2\Phi_S$. This
comparison is shown in \Fref{Fig4}. The error bars in the figure
indicate the accuracy of the fitting procedure with the isochrone and
RABBITT equations \eref{iso} and \eref{RABBITT}, respectively, by
employing the Marquardt-Levenberg nonlinear algorithm.  We observe a
\ask{rather} close agreement between $\Phi_R$ and $2\Phi_S$ in
\Fref{Fig4}. \ask{However, various numerical factors stop us short
  from claiming these two phases leading to the identical atomic time
  delays. Such an identity is established in linear streaking and
  RABBITT \cite{Dahlstrom201353}}. We note that the RABBITT phase
$\Phi_R$ cannot be determined from \Eref{RABBITT} using a single value
of $\t$ as the magnitude coefficients $A,B$ are not known {\em
  ab~initio} and cannot be determined experimentally. Only the whole
set of RABBITT measurements or simulations can be fit by
\Eref{RABBITT} with a common $\Phi_R$ value. Thus RABBITT is not
suitable for a random shot phase determination. In addition, to cover
a wide photoelectron energy range of \Fref{Fig4}, the large harmonic
orders in their 40s should be used in the RABBITT simulations for the
hydrogen and Yukawa atoms. This large order leads to progressively
increasing error bars in the RABBITT data. We note that the typical
harmonic orders in their 20s are employed in most RABBITT
experiments. This observation demonstrates another advantage of the
proposed scheme which can probe large photoelectron energies without
compromising accuracy since the XUV and IR photon energies are
entirely decoupled.

\ask{The phase of the angular integrated RABBITT signal can be converted
to the atomic time delay as $\t_R=\Phi_R/(2\w)$.
Similarly, the streaking phase can be converted to the time delay as
$\t_S=\Phi_S/\w$.} Both conversions are shown on the right axis of
\Fref{Fig4}. As the hydrogen atom is free from many-electron
correlation and resonances, its atomic time delay is known
analytically \cite{Pazourek2013,Serov2015b}. Comparison of both the
ASX and RABBITT time delays with this analytic test is rather accurate
across a broad range of XUV energies. 
As an additional test, we run a set of analogous streaking and RABBITT
calculations on the Yukawa atom in which the Coulomb potential is
screened as $(Z/r)\exp(-r/a)$, where $Z=2.785$ and $a=0.5$, thus
maintaining the hydrogen atom ionization potential. Both the streaking
and RABBITT phases vanish in the Yukawa atom thus indicating the
Coulomb origin of the atomic time delay in hydrogen. Besides solving
the TDSE, we obtain the PMD from the SPM simulations as described in
\cite{PhysRevA.104.033118}.  The streaking phase vanishes once the
same phase extraction method is applied to the SPM data. This provides
additional \ask{credibility} to  the proposed technique.

\begin{figure}[t]
\ms

\hs{-5mm}
\epsfxsize=9.5cm
\epsffile{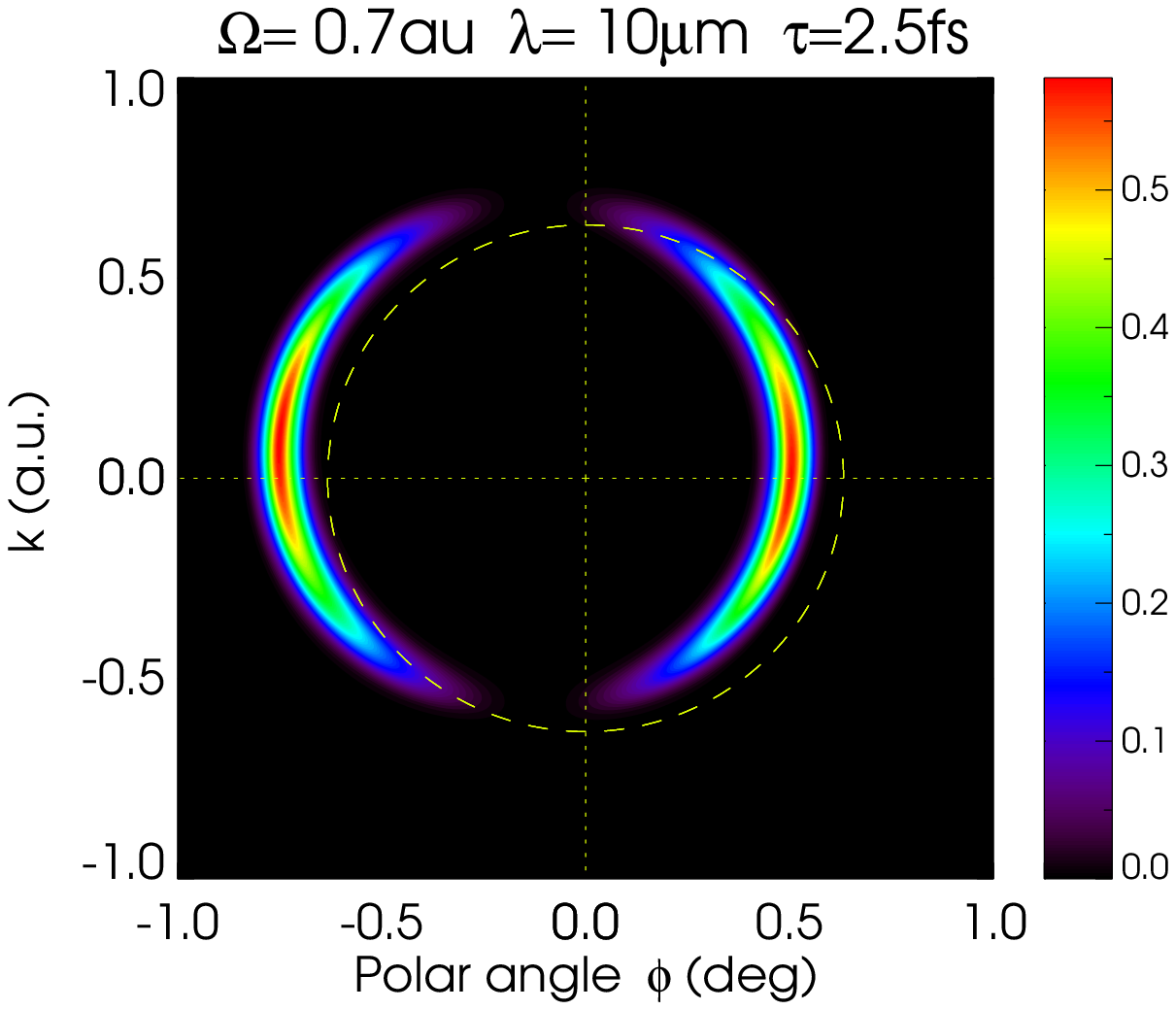}
\ms\ms

\epsfxsize=7cm
\epsffile{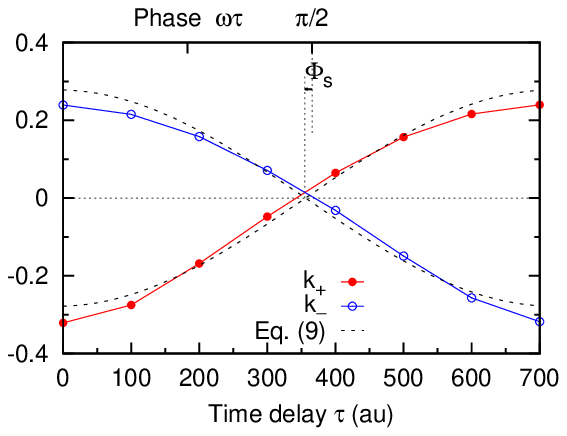}

\mms
\caption{\ask{Top: the PMD of the hydrogen atom in the polarization plane
  at $\Omega=0.7$~au, $\l=10.6~\mu$m and $\tau=2.5$~fs is displayed in
  the Cartesian coordinates. Bottom: radial momentum displacements
  $k_\pm^2/2-k_0^2/2$ are shown at various XUV/IR delays $\t$. The
  dashed line represents the fit with \Eref{ansatz}. The arrow
  indicates the streaking phase $\Phi_S$.}
\label{Fig5}}
\ms
\end{figure}

\ask{Furthermore, we extended our ASX simulations to a different
  regime utilizing a mid-IR laser pulse at $\l=10.6~\mu$m as in the
  experiment \cite{Hartmann2018}. The IR intensity was varied between
  \Wcm{1.5}{9} and \Wcm{2.4}{10} to demonstrate the stability of the
  streaking phase extraction. Results of these simulations at a fixed
  XUV photon energy $\Omega=0.7$~au are shown in \Fref{Fig5}. The top
  panel displays the PMD in the polarization plane at the IR intensity
  of \Wcm{2.4}{10} and the XUV/IR time delay $\tau=2.5$~fs. The PMD
  displays both the radial and angular displacement of the maxima away
  from XUV polarization direction. With the radial displacements being
  determined in these maxima directions, the isochrone analysis is shown
  in the bottom panel of \Fref{Fig5}. While the streaking phase is
  decreased to $\Phi_S=-0.046$~rad, the corresponding time delay grows to
  $\tau_S= 260$~as. This compares with the time delay value
  $\tau_S=125$~as at 1200~nm. Such an increase is expected as
  the CC correction  contains the term 
\be
\tau_{\rm cc}\propto  -{Z\over k^3}
\left[
\ln 1.16 {k^2\over\w}-1
\right]
\ ,
\ee
where $k\simeq0.63$ is the photoelectron momentum and $Z\simeq1$ the
Coulomb charge of the ion remainder \cite{Serov2015b}.  }

\bs
\ask{\section{Conclusions}}
\ms

In conclusion, we demonstrate an accurate retrieval of the streaking
phase $\Phi_S$ from XUV atomic ionization in the presence of a
circularly polarized IR laser field.  The rational of this method is
based on introduction of the streaking phase $\Phi_S$ into the
isochrone equation \eref{iso} derived earlier in \cite{Kazansky2016}
within the SFA. The modified isochrone \eref{isom} has two phase
shifts. One is due to the XUV/IR delay $\w\t$ which can be controlled
by an incremented displacement of the short XUV pulse relative to the
long IR pulse. An additional term $\Phi_S$ is due to the energy
dependent phase of the dipole ionization amplitude \eref{alpha}.
\ask{ Our numerical simulations on hydrogen indicate that such an
  extracted phase can be converted to the atomic time  delay
$\tau_S = \Phi_S/\w$
similarly to the analogous conversion of the phase of the RABBITT
oscillation
$\tau_R = \Phi_R/(2\w) \ .$
Even though the numerical values of $\tau_S$ and $\tau_R$ may differ
insignificantly in our simulations, their similarity across a wide
range of photon energies, right from the ionization threshold, is
quite noticeable}.

Even though the rational of the proposed phase extraction is drawn from
the SFA, its validity is actually much wider and applies to
photoelectrons with the energy as low as only 10\% of the ionization
potential.
Simplicity of the hydrogen atom allows us to test the ASX and RABBITT
time delay $\tau_a$ against the analytical results
\cite{Pazourek2013,Serov2015b}. These results allow to interpret the
atomic time delay as the time it takes for the electron to be
photoionized and the time it takes for the measurement process to
occur. This interpretation was challenged recently for the Coulombic
systems \cite{Saalmann2020} but seems to be confirmed by the present
study. In addition, the proposed technique was extended to the H$_2$
molecule \cite{Serov2022}. Not only did we show an accurate phase
extraction. We also reproduced orientation and two-center interference
effects which are specific to homo-nuclear diatomic molecules.
 
The proposed ASX phase retrieval can be utilized in FEL by collecting
a sequence of XUV radiation shots with random arrival times
$\tau$. This sequence allows to determine the effective vector
potential magnitude $A_0$ from the maximum vertical displacement of
the two PMD lobes. This determination, conducted at a sufficiently
large photoelectron energy, will point to the zero net displacement of
the XUV and IR pulses $\t=0$. This arrival time calibration along with
the known effective $A_0$ will allow to conduct the phase
determination at the whole set of collected data points. Importantly,
the magnitude and  phase of the XUV pulse, which may vary from
shot to shot, do not affect the proposed phase determination as
these parameters do not enter the isochrone equations \eref{iso} and \eref{isom}.

In parallel with the absolute streaking phase determination, one can
use a reference of a second ionization state in the same target or
ionization from an additional, reference target mixed with the
original target, to determine their relative ionization phase. The
reference and target photolines are produced by the same XUV pulse, so
they share a common global streaking angle. Their difference in
streaking angle will give $\tau_a - \tau_r$, where $\tau_r$ is the
reference EWP group delay. If we choose a reference photoline that is
much higher in energy and has a clear continuum, then $\tau_r$ will be
negligible.  This technique has been implemented and analyzed in the
SFA framework \cite{ATTO2019}, where the radial ansatz \Eref{ansatz}
can be used to directly read on a single-shot basis.

VVS was supported by the Australian Research Council Discovery Project
DP190101145.  IAI was supported by the Institute for Basic Science
under the grant IBS-R012-D1. JPC, AM and ALW were supported by the
U.S. Department of Energy (DOE), Office of Science, Office of Basic
Energy Sciences (BES), Chemical Sciences, Geosciences, and Biosciences
Division (CSGB).


\end{document}